\documentstyle[12pt,epsfig]{article}
\begin{document}
\font\ninerm = cmr9
% \baselineskip 28pt plus .5pt minus .5pt
% \pageno=0\footline={\ifnum\pageno>0 \hss --\folio-- \hss \else\fi}

\def\footnoterule{\kern-3pt \hrule width \hsize \kern2.5pt}

\pagestyle{empty}

% \begin{flushright}
% astro-ph/0201012 \\
% August 2000
% \end{flushright}

\vskip 0.5 cm

%\vskip 1.5 cm
\begin{center}
{\large\bf QUANTUM SPACE-TIME:
DEFORMED SYMMETRIES VERSUS BROKEN SYMMETRIES\footnote{Invited talk given at
the 2nd Meeting on CPT and Lorentz Symmetry (CPT 01),
Bloomington, Indiana, 15-18 Aug 2001. To appear in the proceedings.}}
\end{center}
\vskip 1.5 cm
\begin{center}
{\bf Giovanni AMELINO-CAMELIA}
\end{center}
\begin{center}
{\it $^a$Dipart.~Fisica,
Univ.~Roma ``La Sapienza'',
P.le Moro 2, 00185 Roma, Italy}
\end{center}

\vspace{1cm}
\begin{center}
{\bf ABSTRACT}
\end{center}

{\leftskip=0.6in \rightskip=0.6in
Several recent studies have concerned the faith
of classical symmetries in quantum space-time.
In particular, it appears likely that quantum (discretized, noncommutative,...)
versions of Minkowski space-time would not enjoy the classical
Lorentz symmetries.
I compare two interesting cases: the case in which the classical symmetries
are ``broken", {\it i.e.}
at the quantum level some classical symmetries
are lost, and the case in which
the classical symmetries
are ``deformed", {\it i.e.}
the quantum space-time has as many symmetries as its classical
counterpart but the nature of these symmetries is affected
by the space-time quantization procedure.
While some general features, such as the emergence of
deformed dispersion relations, characterize both
the symmetry-breaking case and the symmetry-deformation case,
the two scenarios are also characterized by sharp differences,
even concerning the nature of the new effects predicted.
I illustrate this point within an illustrative calculation
concerning the role of space-time symmetries in the evaluation
of particle-decay amplitudes.
The results of the analysis here reported also show
that the indications obtained by certain dimensional
arguments, such as the ones recently considered
in hep-ph/0106309 may fail to uncover
some key features of quantum space-time symmetries.}

%\pacs{{\tt$\backslash$\string pacs\{\}} }

\newpage

\baselineskip 12pt plus .5pt minus .5pt

\pagenumbering{arabic}
\pagestyle{plain}

\section{Introduction}

In recent years
the problem of establishing what happens to the symmetries
of classical spacetime when the spacetime is quantized has
taken central stage in quantum-gravity research.
In particular, the symmetries of classical flat (Minkowski)
spacetime are well verified experimentally, so it appears that
any deviation from these symmetries that might emerge from
quantum-gravity theories would be subject to severe
experimental constraints. As a result symmetry tests
are a key component of the programme
of ``Quantum-Gravity Phenomenology"~\cite{gacqgplect,rovhisto,carlip}.

In this lecture I focus on the faith of Lorentz invariance
at the quantum-spacetime level. A large research effort has been
devoted to this subject. Most of these studies focus
on the possibility that Lorentz symmetry might be ``broken"
at the quantum level; however, I have recently shown that
Lorentz invariance might be affected by spacetime quantization
in a softer manner: there might be no net loss of symmetries
but the structure of the Lorentz transformations
might be affected by the quantization procedure~\cite{gacdsr,dsr3}.
My primary objective here will be the one of drawing a clear
distinction between the broken-symmetry and the deformed-symmetry
scenarios.

\section{Quantum-Gravity Phenomenology}

Quantum-Gravity Phenomenology~\cite{gacqgplect}
is an intentionally vague name for a new approach to research
on the possible non-classical (quantum) properties of spacetime.
This approach does not adopt any
particular belief concerning the structure of spacetime
at short distances ({\it e.g.}, ``string theory", ``loop quantum gravity"
and ``noncommutative geometry" are seen as equally deserving
mathematical-physics programmes).
It is rather the proposal that quantum-gravity research should
proceed just in the familiar old-fashioned way: through small incremental
steps starting from what we know and combining mathematical-physics
studies with experimental studies to reach deeper and deeper layers
of understanding of the short-distance structure of spacetime.
Somehow research on quantum gravity has wondered off this traditional
strategy: the most popular quantum-gravity approaches, such as
string theory and loop quantum gravity, could be described as ``top-to-bottom
approaches" since they start off with some key assumption about
the structure of spacetime at the Planck scale and then they try
(with limited, vanishingly small, success)
to work their way back to the realm of doable experiments.
With Quantum-Gravity Phenomenology I would like to refer to all
studies that are somehow related with a ``bottom-to-top approach"
to the quantum-gravity problem.

Since the problem at hand is really difficult (arguably the most challenging
problem ever faced by the physics community)
it appears likely that the two complementary
approaches might combine in a useful way: for the ``bottom-to-top approach"
it is important to get some guidance from the (however tentative)
indications emerging
from the ``top-to-bottom approaches", while for ``top-to-bottom approaches"
it might be very useful to be alerted by quantum-gravity phenomenologists
with respect to the type of new effects that could be most stringently
tested experimentally (it is hard for ``top-to-bottom approaches" to
obtain a complete description of low-energy physics, but perhaps it
would be possible to dig out predictions on some specific spacetime features
that appear to have special motivation in light of the corresponding
experimental sensitivities).

Until very recently the idea of a Quantum-Gravity Phenomenology,
and in particular of attempts of identification of experiments with
promising  sensitivity, was very far from the main interests of
quantum-gravity research. One isolated idea had been circulating
from the mid 1980s: it had been realized~\cite{ehns,huetpesk,kostcpt}
that the sensitivity of CPT tests using the neutral-kaon
system has improved to the point that even small effects of
CPT violation originating at the Planck scale\footnote{The possibility
of Planck-scale-induced violations of the CPT symmetry has been
extensively considered in the literature.
One simple point in support of this possibility comes from
the fact that the CPT theorem, which holds in our present conventional
theories, relies on exact locality, whereas in quantum gravity
it appears plausible to assume lack of locality at Planckian scales.}
might in principle be revealed.
These pioneering works on CPT tests were for more than a decade
the only narrow context in which the implications of quantum gravity
were being discussed in relation with experiments,
but over the last 3 years several new ideas for tests
of Planck-scale physics have appeared at increasingly fast pace,
leading me to argue~\cite{gacqgplect} that the times might be right
for a more serious overall effort in Quantum-Gravity Phenomenology.
At the present time there are several examples of experimentally accesible
contexts in which conjectured quantum-gravity effects are being
considered, including studies of in-vacuo dispersion using gamma-ray
astrophysics~\cite{grbgac,billetal},
studies of laser-interferometric limits on quantum-gravity induced
distance fluctuations~\cite{gacgwiFIRST,gacgwiLATEST},
studies of the role of the Planck length in the determination
of the energy-momentum-conservation threshold conditions
for certain particle-physics processes~\cite{kifu,kluz,aus,gactp,jaco},
and studies of the role of the Planck length in the determination
of particle-decay amplitudes~\cite{gacpion}.
These experiments might represent the cornerstones
of quantum-gravity phenomenology since they are as close as one can
get to direct tests of space-time properties, such as space-time
symmetries. Other experimental proposals that should be seen
as part of the quantum-gravity-phenomenology programme rely
on the mediation of some dynamical theory in quantum space-time;
comments on these other proposals can be found in
Refs.~\cite{gacqgplect,veneziano,peri,garaytest,ahlunature,lamer}.

%different levels of QGphenomenology....
%too many parameters....
%few parameters....

The primary challenge of quantum-gravity phenomenology
is the one of establishing the properties of space-time at Planckian
distance scales.
However, there is also recent discussion of the possibility that
quantum-spacetime effects might be stronger than usually expected,
{\it i.e.} with a characteristic energy scale that is much smaller
(perhaps just in the TeV range!) than the Planck energy.
Examples of mechanisms leading to this possibility are found
in string-theory models with large extra dimensions~\cite{led}
and in certain noncommutative-geometry models~\cite{ncstrings}.
The study of the phenomenology of these models of course
is in the spirit of quantum-gravity phenomenology, although
it is of course less challenging than the quantum-gravity-phenomenology
efforts that pertain effects genuinely at the Planck scale.

\section{The faith of Lorentz symmetry in quantum spacetime}

If the Planck length, $L_p$, only has the role we presently attribute
to it, which is basically the role of a coupling constant
(an appropriately rescaled version of the coupling $G$),
no problem arises for FitzGerald-Lorentz contraction,
but if we try to promote $L_p$ to the status of an intrinsic
characteristic of space-time structure (or a characteristic of
the kinematic rules that govern particle propagation in space-time)
it is natural to find conflicts with FitzGerald-Lorentz contraction.

For example, it is very hard (perhaps even impossible)
to construct discretized versions or non-commutative versions
of Minkowski space-time which enjoy ordinary
Lorentz symmetry.
Pedagogical illustrative examples of
this observation have been discussed, {\it e.g.},
in Ref.~\cite{hooftlorentz} for the case of discretization
and in Refs.~\cite{majrue,kpoinap}
for the case of non-commutativity.
Discretization length scales and/or non-commutativity length
scales naturally end up acquiring different values for
different inertial observers, just as one would expect
in light of the mechanism of FitzGerald-Lorentz contraction.

There are also dynamical mechanisms (of the spontaneous symmetry-breaking
type) that can lead to deviations from ordinary Lorentz invariance,
it appears for example that this might be possible in string
field theory~\cite{kosteLORENTZ}.

Departures from ordinary Lorentz invariance are therefore rather
plausible at the quantum-gravity level.
Here I want to emphasize that there are at least two possibilities:
(i) Lorentz invariance is broken and (ii) Lorentz invariance is deformed.

%language....general Lorentz transformations...

\subsection{Deformed Lorentz invariance}

In order to be specific about the differences between
deformed and broken Lorentz invariance let me focus on
the dispersion relation $E(p)$ which will naturally be modified
in either case.
Let me also assume, for the moment, that the deformation
be Planck-length induced: $E^2= m^2 + p^2 + f(p,m;L_p)$.
If the function $f$ is nonvanishing and nontrivial and
the energy-momentum transformation rules are ordinary (the ordinary
Lorentz transformations) then clearly $f$ cannot have the exact
same structure for all inertial observers. In this case one
would speak of an instance in which Lorentz invariance is broken.
If instead $f$ does have the exact
same structure for all inertial observers, then necessarily
the transformations between these observers must be deformed.
In this case one
would speak of an instance in which the Lorentz transformations
are deformed, but Lorentz invariance is preserved (in the deformed sense).

While much work has been devoted to the case in which Lorentz invariance
is actually broken, the possibility that Lorentz invariance might be
deformed was introduced only very recently by this
author~\cite{gacdsr,jurekdsr,michele,dsr3,rossano}.
An example in which all details of the deformed Lorentz symmetry
have been worked out is the one in which one enforces
as an observer-indepedent statement the dispersion relation
\begin{equation}
L_p^{-2}\left(e^{L_p E}
+e^{- L_p E}-2\right)-\vec{p}^2e^{-L_p E}
=m^2
\label{eq:disp}
\end{equation}
In leading (low-energy) order this takes the form
\begin{equation}
E^2 - \vec{p}^2 + L_p E \vec{p}^2
=m^2~.
\label{eq:displead}
\end{equation}
The Lorentz transformations and the energy-momentum conservation
rules are accordingly modified~\cite{dsr3}.

\subsection{Broken Lorentz invariance}

The case of broken Lorentz invariance requires fewer comments
since it is more familiar to the community.
In preparation for the analysis reported in the next Section
it is useful to emphasize that the same dispersion relation
(\ref{eq:displead}), which was shown in Refs.~\cite{gacdsr,dsr3}
to be implementable as an observer-independent dispersion relation
in a deformed-symmetry scenario, can also be considered~\cite{grbgac} as
a characteristic dispersion relation of a broken-symmetry scenario.
In this broken symmetry scenario the dispersion relation (\ref{eq:displead})
would still be valid but only for one ``preferred" class of inertial
observers ({\it e.g.} the natural CMBR frame) and it would be valid
approximately in all frames not highly boosted with respect to
the preferred frame. In highly-boosted frames one might find the same
form of the dispersion relation but with different value of
the deformation scale (different from $L_p$). All this follows
from the fact that in the broken-symmetry scenario the laws
of transformation between inertial observers are unmodified.
Accordingly also energy-momentum conservation rules are unmodified.

Another scenario in which one finds broken Lorentz invariance
is the one of canonical noncommutative spacetime, in which
the dispersion relation is modified (with different deformation
term~\cite{suss,luisa}), but, again, the energy-momentum
Lorentz transformation rules are not modified.

% many types of broken Lorentz invariance
%
%fundamentally broken...really preferred frame...not believable...
%
%spontaneous...
%
%external field...
%
%LoopQG??

\section{Illustrative example: photon-pair pion decay}

In order to render very explicit the differences between
the broken-symmetry and the deformed-symmetry case in this Section
I consider photon-pair pion decay adopting in one case deformed
energy-momentum conservation~\cite{dsr3}, as required by the deformed
Lorentz transformations of the deformed-symmetry case, and
in another case ordinary energy-momentum conservation, as required
by the fact that the Lorentz transformation rules are unmodified
in the broken-symmetry case, but
for both cases I impose {\bf the same}
dispersion relation (\ref{eq:displead}).

In the broken-symmetry case, combining (\ref{eq:displead})
with ordinary energy-momentum conservation rules,
one can establish a relation between
the energy $E_\pi$ of the incoming pion, the opening angle $\theta$
between the
outgoing photons and the energy
$E_\gamma$ of one of the photons
(the energy $E_\gamma'$ of the second photon
is of course not independent; it is given by
the difference between the energy
of the pion and the energy of the first photon):
\begin{eqnarray}
\cos(\theta) &\! = \!& {2 E_\gamma E_\gamma' - m_\pi^2
+ 3 L_p E_\pi E_\gamma E_\gamma'
\over
2 E_\gamma E_\gamma'
+ L_p E_\pi E_\gamma E_\gamma'} ~,
\label{pithresh}
%\end{equation}
\end{eqnarray}
where indeed $E_\gamma' \equiv E_\pi - E_\gamma$.
This relation shows that at high energies (starting at energies of
order $(m_\pi^2/L_p)^{1/3}$) the phase space available to the decay
is anomalously reduced:
for given value of $E_\pi$ certain values of $E_\gamma$
that would normally be accessible to the decay are no longer
accessible (they would require $cos \theta > 1$).

In the deformed-symmetry case one enforces the deformed
conservation rules~\cite{dsr3}
\begin{eqnarray}
E_\pi = E_\gamma + E_\gamma'~,~~~\vec{p}_\pi
= \vec{p}_\gamma + \vec{p}_{\gamma'} + L_p E_\gamma \vec{p}_{\gamma'} ~,
\label{cons}
%\end{equation}
\end{eqnarray}
which, when combined again with (\ref{eq:displead}), give raise
to the different relation
\begin{eqnarray}
\cos(\theta) &\! = \!& {2 E_\gamma E_\gamma' - m_\pi^2
+ 3 L_p E_\gamma^2 E_\gamma' + L_p E_\gamma E_\gamma'^2
\over
2 E_\gamma E_\gamma'
+ 3 L_p E_\gamma^2 E_\gamma' + L_p E_\gamma E_\gamma'^2} ~.
\label{pithreshdef}
%\end{equation}
\end{eqnarray}
Here it is easy to check that for all physically acceptable values
of $E_\gamma$ (given the value of $E_\pi$)
one is never led to consider the paradoxical condition $cos \theta > 1$:
there is no severe implication of the deformed-symmetry case
for the amount of phase space available for the decays
(certainly not at energies around $(m_\pi^2/L_p)^{1/3}$,
possibly at Planckian energies).

\section{Closing remarks}

As shown by the illustrative example of calculation presented in the
preceding Section, the differences between the case in which
Lorentz invariance is broken and the case
in which Lorentz invariance is deformed can be very significant
also quantitatively, concerning the nature and the
magnitude of the effects predicted,
besides being quite clearly significant at the conceptual level.

The calculation in the preceding Section also shows that simple
dimensional estimates of the effects induced by deviations from Lorentz
invariance are futile. Both in the broken-symmetry case and in the
deformed-symmetry case the Planck-scale deformation introduces
the same correction terms, respectively in Eqs.~(\ref{pithresh})
and (\ref{pithreshdef}), but in the broken-symmetry case I found
profound implications for pion decay, whereas in the deformed-symmetry
case the correction terms arranged themselves in a less ``armful"
manner.
This observation appears to be particularly significant for the
argument recently presented by
Brustein, Eichler and Foffa in Ref.~\cite{brust}: in that paper,
by applying dimensional analysis to some aspects of neutrino physics
it was suggested that Planck-scale-induced deviations from ordinary
Lorentz invariance are unlikely. The fact that the analysis reported
in Ref.~\cite{brust} relies on the type of dimensional arguments
which I have here shown to be inclusive, forces us to assume
that the conclusions drawn in Ref.~\cite{brust} are equally unreliable.
Certainly the observations reported in Ref.~\cite{brust} provide
strong motivations for future dedicated and rigorously quantitative
studies of the relevant aspects of neutrino physics within specific
examples of Planck-scale-induced deviations from ordinary
Lorentz invariance.

\section*{Acknowledgments}
I am grateful for the encouraging feed-back received from several
participants to CPT01, and particularly for conversations
with Roman Jackiw and
Alan Kostelecky.

%\section*{References}

\end{document}